\documentstyle[12pt]{article}

\begin{document}
\bibliographystyle{unsrt}

\begin{flushright} UMD-PP-95-21

\today
\end{flushright}

\vspace{6mm}

\begin{center}

{\Large \bf Constraints on Massive Tau Neutrinos and Their Cosmological
Implications}\\ [6mm]

\vspace{6mm}
{\bf{R.N. Mohapatra\footnote{Work supported by the
 National Science Foundation Grant PHY-9119745}~and~
S. Nussinov\footnote{Permanent address: Department of Physics and
Astronomy, Tel Aviv University, Tel Aviv, Israel.}}}

{\it{ Department of Physics and Astronomy,}}
{\it{University of Maryland,}}
{\it{ College Park, MD 20742 }}

\end{center}
\vspace{20mm}
\begin{center}
{\bf Abstract}
\end{center}
\vspace{6mm}

We point out the following astrophysical consequences
of a tau neutrino with mass in the MeV range;
(i) if it has a small electric charge which will allow
 it to become a cold dark matter of the universe, then
 present limits on the 511 KeV gamma ray line
rule out the possibility that it  contributes an
 $\Omega_{\nu_{\tau}}$ between .1 to 1
making it unsuitable as a  cold dark matter candidate;
 (ii) if an electrically neutral MeV range $\nu_{\tau}$ decays
to $\nu_e + \chi$ ( where $\chi$ is a massless boson) , then its lifetime
is bounded by SN1987A observations to be within a window $.05~~sec.\leq
\tau(\nu_{\tau})~\leq 300~sec./(m_{\nu_{\tau}}~in~ MeV)$ a range which
may be of interest from the point of view of structure formation.
Encouraged by the overlap between the range allowed in (ii) above
and that required for a proposed structure formation mechanism
by Dodelson et al., we search for simple extensions of the standard model,
where such masses and lifetimes may arise for natural values of parameters
and show that  the  already existing singlet majoron and low scale
left-right models have this property. We also comment on possible
familon models for this decay.

\newpage

\noindent{\bf I. Introduction:}

\vspace{6mm}

Massive Tau neutrinos with masses in the  range $1~MeV\leq m_{\nu_{\tau}}
\leq 31~ MeV$ are allowed by the present laboratory measurements \cite{wein}
and are not in conflict with cosmological  mass density bounds
\cite{kolb} and structure formation
  provided there is some mechanism ( involving
either decay or annihilation ) that reduces their concentration in the
early universe to an appropriate level. There are also interesting
allowed ranges of lifetimes of MeV range $\nu_{\tau}$'s which maintain
the predictions for primordial light element abundances in agreement
with observations\cite{gary}.  It has recently been suggested that
one may be able to use such neutrinos for
some positive purposes such as being the cold dark matter of the universe
\cite{fl} or for helping in understanding the nature of large scale
structure in the universe\cite{dgt}. In this paper, we take a closer look
at these suggestions to see if they are consistent with some other
 astrophysical considerations.

One suggestion is that if the $\nu_{\tau}$ is a Dirac particle with mass
in the  MeV range and has
  an anomalous electric charge\cite{fl}
(in the range $3\times 10^{-6}~\leq ~Q_{\nu_{\tau}}~\leq 3\times 10^{-5}$ )
, then its concentration in the early universe can be
reduced via the annihilation
process $\nu_\tau \bar{\nu}_{\tau}\rightarrow e^+ e^-$ leaving a
freezeout residual density smaller than the cosmological bound. The stable
remnant $\nu_\tau$'s then could serve as the dark matter on cosmological
and galactic scales. Note that the supernova bounds on the Dirac neutrino
masses\cite{gandhi} do not apply in this case since the right-handed neutrinos
having "strong" electromagnetic interactions get trapped in the supernova
after their production.

A second suggestion due to Dodelson, Gyuk and Turner\cite{dgt}
 is that a $\nu_{\tau}$
in the above mass range which decays as $\nu_{\tau}\rightarrow \nu_e~+~\chi$,
where $\chi$ is a mass less boson ( say the majoron or the familon )
with lifetime in the range of $0.1 ~sec\leq \tau_{\nu_{\tau}}~\leq 10^3~sec$
enhances the fraction of energy in the relativistic components in the universe
at the time of its decay. This has the effect of delaying the onset of
the perturbation growth and structure formation at the corresponding
horizon scales. This in turn allows a better over all fit to a pure
cold dark matter model scenario to the fluctuation power spectrum deduced
from the study of galactic redshifts and from COBE\cite{mather} .
An important aspect of this scenario is that the $\nu_e$'s produced in this
decay can destroy the otherwise overproduced $^4$He ( and
other light elements )
and for a certain range of $\tau$ and $m_{\nu_{\tau}}$
one can achieve an acceptable
big bang nucleosynthesis\cite{gary,gyuk}.

The properties of $\nu_{\tau}$ invoked in the above two cases were motivated
by different reasons but they are none-the-less of physical interest from
the point of view of current experiments\cite{gomes} where there is a
possibility that the upper limit on the tau neutrino mass can be reduced
to the level of 3-5 MeV. This will of course provide the ultimate test
of these proposals. Our goal however is to seek other constraints on these
scenarios in order to test their viability. We find that :

(i) The present limits on the 511 keV gamma rays that result from $e^+e^-$
annihilation of any massive charged dark matter would imply that the
contribution of these dark matter particles to $\Omega$ will at most be
ten percent, thereby making such particles not viable as potential dark matter
candidates. These arguments apply not just to tau neutrinos but to any massive
minicharged particle which is a potential dark matter candidate.

(ii)We then show that the time profile of the observed neutrino pulses from
SN1987A at Kamiokande and IMB implies that for a decay $\nu_{\tau}
\rightarrow \nu_i~+~\chi$ ( where i = e or $\mu$ ) with a sterile $\chi$,
the $\nu_{\tau}$ lifetime has a lower bound
$$\tau_{\nu_{\tau}}~\geq~0.05~sec\eqno(1)$$
Moreover, if in the final state of the decay process, the
 $\nu_e$ mode dominates
over the $\nu_{\mu}$ mode, then one can also derive an upper bound :
$$\tau(~\nu_{\tau}\rightarrow \nu_e + \chi~)~\leq 30-{300}~sec.~for~
{m_{\nu_{\tau}}~=~10~ to~1~MeV} \eqno(2)$$.

It is amusing that there is a considerable overlap between the range defined
by our bounds and the one preferred by Dodelson et al. from the structure
formation considerations\cite{dgt}. Encouraged by this, we proceed to
note that two existing simple extensions of the standard model
( the singlet majoron model\cite{cmp} and the minimal low scale left-right
symmetric model\cite{ms} ) can
naturally provide tau neutrino masses and lifetimes in this range. This
should provide additional stimulus for taking this scenario for structure
formation seriously and bring new urgency to improving the upper limits on the
tau neutrino mass. In  particular, one model, the minimal singlet majoron
model\cite{cmp} remains viable in this context only if the electron
neutrino mass is no less than an eV or so. This can also be tested
in the ongoing neutrinoless double beta decay experiments\cite{klap}.

\vspace{6mm}

\noindent{\bf II. Minicharged tau neutrino as a potential
dark matter candidate:}

\vspace{6mm}

 Before entering into the implications of the minicharged tau neutrino,
let us first remind the reader that although unconventional, there is
no fundamental reason that can prevent us from entertaining the possibility
that  the tau neutrino can have
an electric charge as large as $\simeq 10^{-5}$ ( in units of the positron
charge)\cite{foot}. We recall that SN1987A observations on the other hand
imply that the charge on
the electron neutrino cannot be larger than $\simeq~10^{-17}$\cite{barb} .
 Therefore, any theoretical model that is constructed to lead an
anomalous charge for $\nu_{\tau}$ should respect this stringent bound
on $Q_{\nu_e}$. One way to achieve this is
 to gauge an anomaly free quantum number
$L_\mu-L_\tau$ in the standard model and have an elctric charge formula
of the type\cite{bv}:
$$Q~=~I_{3L}~+~{{Y}\over{2}}~+~\epsilon~(L_\mu-L_\tau)  \eqno(3)$$

The electron neutrino in this model is electrically neutral whereas
the $\nu_{\mu}$ and $\nu_{\tau}$ have equal and opposite charges.

 Let us now consider the minicharged tau neutrino as the
dark matter of the universe and discuss its implications. If the
charge on the tau neutrino is less than $10^{-5}$ or so, its annihilation
cross-section will go out of equilibrium below the nucleosynthesis
temperature of one MeV and will not effect the remnant cosmological
tau neutrino density. However, the $\nu_{\tau}$'s in the galactic
disc is a continuous source of positrons. These positrons of energy
$E_{e^+}\simeq 1~-~10~MeV$ will be confined by the galactic magnetic fields
inside the disc. Their collisions with the disc gas will slow them down on
time scales much less than the age of the universe and practically
all positrons
will then annihilate at rest. In one fourth of these annihilations, namely
those proceeding from the singlet spin state, two monochromatic $\gamma$-
ray lines of energy 511 keV will emerge. We can estimate the flux of
these $\gamma$-rays locally to be roughly:
$$\Phi_\gamma\approx~[{{1}\over{4}}n^2_{\nu}\sigma^{\nu\bar{\nu}\rightarrow
e^+e^-}.~v].R_{eff}\eqno(4)$$
$R_{eff}$ is an effective distance parameter related to the radius
of the galactic disk and is estimated below.
Taking the local $\nu_{\tau}$ density to be that of the local dark matter i.e.
400 MeV/$(cm)^3$, we have
$$n_{\nu_{\tau}}~\simeq~\left({{400}\over{m_{\nu_{\tau}}/MeV}}\right)(cm^{-3})
\eqno(5)$$
The $\nu_{\tau}\bar{\nu}_{\tau}\rightarrow e^+e^- $ annihilation cross-section
for $2~m_{\nu_{\tau}}$ sufficiently above the threshold can be written as:
$$\sigma.(v/c)\simeq {{2\pi\delta^2_{\tau}\alpha^2}\over{4~m^2_{\nu_{\tau}}}}
\simeq~ 3\times 10^{-36}cm^2\left({{\delta_\tau}\over{10^{-5}}}\right)^2
\left({{MeV}\over{m_{\nu_{\tau}}}}\right)^2\eqno(6)$$
For a cylinder like geometry of the galactic disc with a radius $R\simeq 20$
kilopersecs  and thickness $D\approx 1$ kilopersec, the effective
radius $R_{eff}$
in eq.(4) can be estimated to be roughly $\approx (R/ 4\pi)$, leading
to a value for $R_{eff}\approx 1.5$ kpersecs. Substituting back into eq.(4),
we get for the 511 keV gamma ray flux generated by this mechanism to be:
$$\Phi_{\gamma}\simeq 20({{\delta}\over{10^{-5}}})^2.({{MeV}\over
{m_{\nu_\tau}}})^4~cm^{-2}~sec.^{-1}~sr^{-1}\eqno(7)$$
In ref\cite{fl}, it was shown that the prediction for $\Omega_{\nu_{\tau}}$
in this model is given by,
$$\Omega_{\nu_{\tau}}\approx {{1}\over{10}}({{\delta}\over{10^{-5}}})^{-2}
({{m_{\nu_{\tau}}}\over{MeV}})^2\eqno(8)$$
For $\Omega_{\nu_{\tau}}\leq~1$  eq.(7) and (8) imply that,
$$\Phi_{\gamma}\geq~2\left({{MeV}\over{m_{\nu_{\tau}}}}\right)^2~cm^{-2}
{}~sec.^{-1}~sr^{-1}\eqno(9)$$
For tau neutrino masses less than 10 MeV, eq.(9) implies a lower bound on
the 511 keV photon flux more than $\approx 2\times 10^{-2} cm^{-2}~sec^{-1}
{}~sr^{-1}$.
The experimental study of gamma rays and the 511 keV line in particular
has been an ongoing effort for many years. A  fit\cite{ramaty}
 to the recent GRO\cite{gro} data implies a flux of 511 keV line in the
disc  (in the direction away from the galactic center) at
the level of $5\times 10^{-4}~cm^{-2}~sec^{-1}~sr^{-1}$
 and would therefore
strongly argue against the minicharged tau neutrino scenario for dark matter.

 It has already been noted in \cite{bv} that there exist
 independent theoretical arguments  against a
 tau neutrino charge larger than about $10^{-6}$ if one uses eq.(3).
As already mentioned, eq.(3)
implies that  the muon neutrino and the
tau neutrino must have equal and opposite charges in order to satisfy
anomaly cancellations. The existing data on
 $\nu_{\mu}e$ scattering\cite{abe} can then be used to set an
upper limit of about $10^{-9}$ on the electric charge of the muon neutrino
and hence by eq.(3) on $\delta_{\nu_{\tau}}$. Moreover,electric charge
conservation in muon decay implies
 that the muon also must have slight anomalous
charge equal to $\delta_{\nu_{\tau}}$. The precision measurements involving
the muon then imply severe restrictions on $\delta_{\nu_{\tau}}$.
For instance, one gets\cite{bv}
$|\delta_{\nu_{\tau}}|\leq 10^{-6}$  from the measured value of
 $g-2$ of the muon\footnote{ One could also conceivably improve on this bound
by a dedicated experiment where one could look for the displacement of
the center of mass of the muonium  atom in  strong electric field.}.
 These considerations also independently exclude the minicharged tau neutrino
scenario for cold dark matter,
 when it is embedded into the gauge model framework.

\newpage
\noindent{\bf III. SN1987A limits on the $\nu_{\tau}$ life time:}
\vspace{6mm}

The cooling of the hot compact cores of the stars  following
gravitational stellar collapse ( eventually leading to a type II supernova )
is believed to occur via neutrino emission. The observed neutrino
pulses from the SN1987A \cite{hirata} beautifully confirm this
expectation in several respects:

(i). The overall $\bar{\nu}_e$ fluence is consistent with a total
collapse energy of 3-5$\times~10^{53}$ ergs that can be associated with
the gravitational binding energy of the star.  This energy is
released in roughly equal amounts via each of the six neutrino
species: $\nu_e,\nu_{\mu}, \nu_{\tau}$ and their anti-particles.

(ii). The energy spectrum of the observed $\bar{\nu}_e$'s is consistent
with an expected thermal spectrum with temperature$\approx~4~MeV$.

(iii). The duration of the pulse $\Delta t\approx$ 5-12 sec. can be
interpreted as the "trapping" time of the initial energetic neutrinos
in the dense core. The neutrinos diffuse gradually into the less dense
regions and also "cool off" in the process. The effectiveness of the trapping
which is governed by
$\simeq n_{e,p,n}\times \sigma_\nu$
 diminishes as the number density of electrons and nucleons
 $n_{e,p,n}$ goes down simultaneously with the fact that $\sigma_{\nu}$
also goes down as the neutrino energy  decreases. As a result,
 the neutrinos escape after a time
$\Delta t$. The demand that no additional, equally efficient cooling
mechanism exists so that the above successful predictions and in particular
point (iii) will not be lost, leads to new bounds on interactions that
provide new cooling mechanisms.

The two body decay with the sterile massless boson $\chi$ ( majoron or
familon ) mentioned
earlier provides one such cooling mechanism since the massless boson
$\chi$ escapes as soon as it is produced\footnote{In principle, the emitted
$\chi$'s can be captured on the ambient $\nu_e$'s and reform $\nu_{\tau}$'s
as "resonances". The relevant cross-section at the peak of the resonance is
$\sigma_{res}\approx {{4\pi}\over{m^2_{\nu_{\tau}}}}$. However, the probability
for having in the $\nu_e \chi$ collision exactly the $E_{cm}= m_{\nu_{\tau}}
\pm {\Gamma/2}$ with $\Gamma\approx (sec.)^{-1}\simeq 6\times 10^{-22}~MeV$
is very small. The back reaction rate is proportional to an effective
cross-section $\sigma_{eff}\simeq \sigma_{res}{{\Gamma}\over{E_{\nu}}}\approx
10^{-45}cm^2$ for $E_\nu\simeq m_{\nu_{\tau}}\simeq 10~MeV$. This value of
$\sigma_{eff}$ is much smaller than the ordinary weak interaction
cross-section that traps neutrinos.} .
 If the decay time is much shorter
the above $\Delta t$ i. e.( denoting
$\tau_{\nu_{\tau}}\equiv \tau$ )
$\tau.\gamma\equiv \tau(E_{\nu_{\tau}}/m_{\nu_{\tau}})\ll \Delta t\simeq
5-12 ~sec$, then this is indeed an efficient cooling mechanism.
Demanding that this not be the case yields a lower bound on $\tau$ of about
.1 sec or so.

Conversely, if $\nu_{\tau}$ of mass about 10 MeV decays predominantly into
$\nu_e~+~\chi$ with a relatively long lifetime($\geq$100 - 1000 sec.), then
the decay $\nu_e$ would lead to a delayed pulse in the underground detector
beyond $\Delta t\simeq 5~-~12~sec.$. This reasoning parallels the the one made
in connection with the 17 keV neutrino\cite{mn} yields an upper bound on
 $\tau$.

Let us now elaborate on both these arguments . First we discuss the
derivation of the lower bound. During $\delta t~=~\tau.\gamma~=~\tau(E/m)$,
half the energy in the $\nu_{\tau}~+~\bar{\nu}_{\tau}$ components would
 escape the core due to the decay to $\chi$. This energy can be written as:
$$W_{\chi}~=~{{1}\over{2}}(W_{\nu_{\tau}}~+~W_{\bar{\nu}_{\tau}})~=~
{{1}\over{6}}f_B~W_{tot}\eqno(10)$$

In eq.(10), $f_B$ is the Boltzman factor, which is $\simeq 1$ for low
masses of the tau neutrino ( e. g. $m\simeq 1$ MeV) since the tau neutrino
temperature is about 5 MeV. For $m_{\nu_{\tau}}\simeq 10 $MeV, we can
approximate $f_B\simeq ({{m}\over{T}})^{{{3}\over{2}}}e^{-{{m}\over{T}}}
\approx~.36$. Since the thermal equilibrium population of the tau
neutrinos will be immediately regenerated inside the neutrino sphere, the
$W_{tot}$ stored in the core will decrease decrease by a factor $e$ over
a time
$$t^{\chi}(cooling)~=~ (6/f_B)\times (\tau.E/m)\eqno(11)$$
Using $E=4T_{\nu_{\tau}}\simeq 20 MeV$, we get $t^{\chi}\simeq 120\tau$
for a one Mev tau neutrino whereas $t^{\chi}\simeq 40\tau$ for a 10 Mev
mass. In order to make this extra cooling mechanism essentially ineffective,
we demand that it be bigger than about 12 sec. This leads to to the lower
bound on $\tau $ of .1 sec as mentioned above.

Turning now to the upper bound, we noted earlier\cite{mn} that if a
$\nu_{\tau}$ of mass $m$ decays via a two body mode such as $\nu_e~+~\chi$
(with both these particles massless) with a lifetime $\tau$, then
 the $\nu_e$'s arrive earth after a time
$$\delta t~=~({{m}\over{2E_{\nu_{e}}}})\tau\eqno(12)$$
where $E_{\nu_e}$ is the energy of the electron neutrino.
Since on the average,
 $2E_{\nu_e}~=~E_{\nu_{\tau}}$ = 20 MeV, we have for $m= 1~-~10~MeV$,
$\delta t = (.05~-~.5)\tau$. These delayed $\nu_e$'s being roughly
of similar energy to the original neutrinos will generate a delayed
neutrino pulse with roughly 4-13 events in the IMB and 6-19 events
in the Kamiokande detector\cite{mn}. Since  neither detector found any
delayed pulse during the time interval of 15 sec. to an hour after the initial
pulse, eq.(12) implies an upper bound on $\tau$ of about 300 sec. for
$m_{\nu_{\tau}} = 1 MeV$ and 30 sec. for $m_{\nu_{\tau}} = 10 MeV$.
As the tau neutrino mass increases, there is suppression of the emitted
number of tau neutrinos by the appropriate Boltzman factor and that
reduces the overall magnitude of the delayed signal . This for instance
implies that the overall quality of  our bound for a 10 MeV $\nu_{\tau}$
is somewhat weaker than for a 1 MeV $\nu_{\tau}$. Using the estimates
carried out in the second paper of ref.\cite{mn}, we would expect
10-32 delayed events in both experiments for $m_{\nu_{\tau}}= 1~ MeV$
as compared to 3.5-11 events for $m_{\nu_{\tau}}= 10~MeV$.

We wish to point out that the lower
bound does not apply if the final state of the tau neutrino decay involves
three $\nu_e$'s , since unlike the $\chi$'s, the $\nu_e$'s will
get trapped in the supernova due to their weak interactions and no
dramatic cooling can be expected. As for the  delayed neutrino pulse,
on the average, the final state $\nu_e$'s will have one third of the
$\nu_{\tau}$ energy and since in the detector the $\nu_e$ cross-section
goes like $\approx E^2_{\nu_e}$, the secondary delayed pulse will
 be too weak to yield an upper bound on $\tau_{\nu_{\tau}}$

Finally, in closing this section, we note that the SN1987A favored range
for the lifetimes $~.1~sec.\leq\tau\leq~300~sec.$, coincides with that
favored by Dodelson et al in their decaying $\nu_{\tau}$ CDM scenario.
Encouraged by this fortuitous coincidence, we explore in the next section
the question of accomodating such lifetimes in simple particle physics
models.
\vspace{6mm}

\noindent{\bf IV. Gauge models with desired $\nu_{\tau}$ mass and
lifetime:}

\vspace{6mm}
In this section, we consider simple extensions of the standard model
which yield $\nu_{\tau}$ mass in the MeV range and lifetime in the
range of 1 - 100 sec. motivated by the discussions of the previous
sections.
 We will consider see-saw type models which provide the
simplest way to understand the smallness of neutrino masses. Ignoring
small mixing angles between different generations in the first approximation,
the generic mass formula for the light neutrinos in these models
has the form:
$$m_{\nu_{i}}\simeq\left({{m^2_{\nu^D_i}}\over{f_i v_{BL}}}\right)\eqno(13)$$

In eq.(13), $m_{\nu^D_i}$ are the Dirac masses of the neutrinos;
$v_{BL}$ is the scale of $B-L$ symmetry breaking and $f_i$ are
the yukawa coupling of the right-handed neutrinos .In order
to estimate the neutrino masses in such models, an additional input
about the magnitude of these Dirac masses is needed. Only in grand
unified theories, they can be related to the charged fermion masses;
 but in near electro-weak TeV scale theories that
we will be interested in here, they are free parameters.
However, motivated by eventual possible grand unification
it is conventional to assume that even in TeV scale theories, the
neutrino dirac masses are  of the same order
of magnitude as the chaged lepton masses of the corresponding
generation.  If we then assume that $v_{BL}\simeq 100~GeV$ and
$f_i\simeq 1$, one  gets the following order of magnitude
estimates for the different neutrino masses; i.e.
 $\nu_{\tau}$ mass of order $\simeq few~MeV$;
 $\nu_\mu$ mass, about a 100 keV and $\nu_e$
mass , a few eV. In order to satisfy the cosmological constraints,
the model must provide a mechanism for the $\nu_{\tau}$ and $\nu_{\mu}$
to decay sufficiently fast. We will see below that in the two
simplest TeV scale see-saw models , although there are mechanisms
 for them to decay; the decay rate for $\nu_{\mu}$ is much too slow to satisfy
the lifetime bounds coming from structure formation\cite{gs} although
it can easily satisfy the mass density constraints. Since one of
our main motivations for considering this scenario is structure formation,
we will take the constraints implied by it  seriously. This then implies
that the muon neutrino mass must also be in the eV range and for a Tev scale
see-saw this requires $m_{\nu^D_{\mu}}\simeq m_{\nu^D_{e}}\simeq m_e$.
Such a situation could arise if there is some kind of family symmetry
between the first and the second generation.

A second point we wish to make concerns the familon models\cite{gny}
 for $\nu_{\tau}$
decay . For a global family symmetry that operates on the third
generation, present limits on $\tau\rightarrow e~+~f$ imply a lower limit
on the scale of family symmetry breaking, $v_F \geq 10^6 GeV$ or so\cite{pdg}.
If we identify the scale $v_F$ with the $B-L$ symmetry breaking scale
$v_{BL}$, ( as one would be tempted to do in simple models ),
then the see-saw formula produces only a keV range mass for
the $\nu_{\tau}$. Therefore, any familon model for an MeV $\nu_{\tau}$
must decouple the $B-L$ and the Family symmetry breaking scales.

Let us now to proceed to analyze the $\nu_{\tau}$ lifetimes in the
singlet majoron and the low scale left-right models.

\vspace{6mm}
\noindent{\it A. The singlet majoron model:}
\vspace{6mm}

The singlet majoron model\cite{cmp} is the most minimal extension of
 the standard model with nonvanishing neutrino masses and the see-saw
mechanism. In addition to the particles of the standard model it
consists of three right-handed neutrinos , one for each generation
and an additional electroweak singlet Higgs boson $\Delta^0$
that carries total lepton
number,$L~=~2$. The Lagrangian of the model is assumed to obey the
total lepton number symmetry which is spontaneously broken by the
$\Delta^0$ field acquiring a vacuum expectation value(vev) at a scale
around 100 GeV to a TeV. The neutrino mass matrix then has the see-saw
form given in eq.(13) and the earlier discussion about the neutrino
spectrum applies. Let us therefore go on to discuss neutrino decays
in this model.

The possible relevance of the majoron decay modes of heavy neutrinos
for cosmology has been considered in several papers\cite{cmp2}. In particular,
 using the results of the last paper in ref.\cite{cmp2}, we get for
$\tau_{\nu{\tau}}$
$$\tau^{-1}_{\nu{\tau}}~\simeq~sin^22\beta{{m_{\nu_e}~m^4_{\nu_{\tau}}}\over
{16\pi v^4_{BL}}}\eqno(14)$$

In eq.(14), $\beta$ is a mixing angle different from the mixing angles
measured in the neutrino oscillation experiments. We will choose $\beta
= \pi/4$ in order to maximize the decay width in eq. (14).
We see from eq. (14) that for a 10 MeV $\nu_\tau$, allowing the $\nu_e$
mass of about 2 eV to be consistent with neutrinoless double beta decay
bounds, we get  $\tau_{\nu_{\tau}}\simeq 150 ~sec.$, which is at  the edge of
the allowed range of ref.\cite{dgt}. As $\nu_\tau$ mass increases, the lifetime
can get shorter and move well into the allowed range depending on the $\nu_e$
mass. However due to quartic power dependence on the $\nu_{\tau}$ mass,
values of $m_{\nu_{\tau}}$ smaller than 10 Mev will lead to lifetimes
outside the desired range. This is important since it is expected that
in the B-factory, tau neutrino masses down to a few MeV can be probed
\cite{gomes}.Similarly, the mass of the $\nu_e$ must not be too much
smaller than an eV or so if this model is to remain viable. The model
is therefore testable in near future.

\vspace{6mm}

\noindent{\it B. The left-right symmetric model:}

\vspace{6mm}

Let us now discuss the $\nu_{\tau}$ lifetime in
the minimal left-right symmetric model with
a see-saw mechanism for neutrino masses \cite{ms}.
 Let us remind the reader about the leptonic and Higgs sector of the model.
The three generations of
 lepton fields are $\Psi_a \equiv {\pmatrix{\nu \cr e \cr }}_a$,
where $a~ =~ 1,~ 2,~ 3$.
 Under the
gauge group $SU(2)_L \times SU(2)_R \times U(1)_{B-L}$, they transform as
$\Psi_{a~L} \equiv (1/2, ~ 0 , ~ -1 )$
and $\Psi_{a~R} \equiv (0, ~ 1/2, ~ -1 )$.
 The Higgs sector
of the model consists
of the bi-doublet field
$\phi \equiv (1/2, ~ 1/2, ~ 0)$ and triplet Higgs fields:
${
\Delta_L ( 1, ~0, ~ +2 ) \oplus \Delta_R (0, ~ 1, ~ +2 )
{}~~~~~.}$

The gauge symmetry is spontaneously broken by the vacuum expectation
values:${< {\Delta_R^0} > = V_R ~~; }$ ,
${< \Delta_L^0 > =  0 ~~;}$
 and ${< \phi > = \pmatrix{ \kappa & 0 \cr
                    0 & \kappa^\prime \cr } ~~}$.
 As usual, $< \phi >$ gives masses to the charged fermions and Dirac masses
to the neutrinos whereas
$< \Delta_R^0 >$
 leads to the see-saw mechanism for the neutrinos in the standard way
and again the discussion about the neutrino spectrum in the introduction
to this section applies.

The interactions resposible for neutrino decay in this model comes
  from the left-handed triplet sector of the theory.
As indicated above, these fields do not take part in the Higgs mechanism.
 The Yukawa Lagrangian relevant
to our discussions is given ( in the basis where all the leptons are mass
 eigenstates ) by

$${{
{\cal L}_Y =  \nu_L^{T} F^{\prime} C^{-1} \nu_L \Delta^{0}
                  - {\sqrt 2} \nu_L^T F^{''}C^{-1}E_L \Delta^{+}_L
              - E_L^T F C^{-1} E_L \Delta^{++} + h.c ~~~~~~,
                 }
}\eqno(15)$$

 where
$\nu = (
\nu_e, ~
 \nu_\mu, ~ \nu_\tau ),~ E = ( e, ~\mu , ~ \tau ); ~F, ~ F^\prime ~
{\rm and} ~ F^{''}$ are
$3\times 3$ matrices related to each other
as follows:
$${
 K F = F^{''} ; ~~ K F K^T = F^\prime ~~~,
}\eqno(16)$$
 where $K$ is the leptonic CKM
 matrix in the left-handed
 sector. First, we note that exchange of the $\Delta^0$ enables the
heavier neutrinos to decay to the lighter ones\cite{ronc}; secondly,
 the off-diagonal elements of
$K$ are the
$\nu_e ~\nu_\mu, ~~~\nu_e ~\nu_\tau~~~ {\rm and} ~~
\nu_\mu ~ \nu_\tau$ mixing angles, which
are directly
measurable parameters, and are at present restricted to be,
 $\theta_{\mu\tau} \leq 0.03, ~~ \theta_{e \mu} \leq .03
{}~~{\rm and} ~~\theta_{e \tau} \leq .17$\cite{boehm}.
 Furthermore, the present upper
limits on the $\mu \rightarrow 3 ~e$ decay can be satisfied by demanding
$F_{\mu e} \simeq 0$, since otherwise
 a large non-zero $\mu \rightarrow 3 ~e$ decay
could arise via $\Delta^{++}_L$ exchange. Moreover, in several recent papers,
\cite{mnz} it has been pointed out that there are strong upper limits on the
decay mode $\nu_{\tau}\rightarrow e^+e^-\nu_e$ from SN1987A observations.
In order for the left-right model to be consistent with this bound, we
must set $F_{\tau e} = 0$ . This leaves only one off-diagonal coupling
$F_{\tau \mu}$ free. Present upper limits on the $\tau\rightarrow \mu~+~\gamma$
decay can be staisfied for $F_{\tau \mu}\simeq 10^{-2}$ or so.

Before discussing the expectations for $\nu_{\tau}$ lifetime
we wish to note an important property of the model, which is that
the decay amplitudes for $\tau\rightarrow {\mu}ee$ and
 $\nu_{\tau}\rightarrow \nu_{\mu} \nu_e \nu_e$ are not related as can be
seen from eq.(15). In fact,even in the limit of vanishing amplitude for
$\tau\rightarrow \mu ee$ decay ( assured by setting $F_{\tau \mu}=0$),
 the  $\nu_{\tau}$ decay can proceed via the mixing angle $K_{\tau \mu}$.
The implication of this observation is that we can obtain a desired
value for $\tau_{\nu_{\tau}}$ without running into conflict with
the upper limits on the rare decays of the $\tau$-lepton.

Let us now discuss this in detail. Note that the lifetime for
the decay  $\nu_{\tau}\rightarrow \nu_{\mu}\nu_e\nu_e$ is given by:
$$\tau^{-1}_{\nu_{\tau}\rightarrow \nu_{\mu} \nu_e \nu_e}~=~
{{{F^{\prime}}^2_{\tau \mu} F^2_{ee}m^5_{\nu_{\tau}}}\over
{768\pi^3 M^4_{\Delta^0}}}\eqno(17)$$

If we choose $F_{\tau \mu}\simeq =0$ in order to forbid the rare
$\mu ee$ decay of the tau lepton, then in eq. (17)
$F^{\prime}_{\tau \mu}\simeq \theta_{\tau \mu}\times (F_{\tau \tau}-
F_{\mu \mu})$. Using the present limits on the neutrino mixing
angles, we can have $F^{\prime}_{\tau \mu}$ of order $0.06$.
Assuming $F_{ee}\simeq .1 -1 $, we then get
tau neutrino lifetime of about .2 to 20 sec. which is in the range desired
in \cite{dgt}. Only if the upper limit on $\theta_{\tau \mu}$ goes below
$10^{-3}$, then one has to invoke a non-zero $F_{\tau \mu}$ ( and
hence a nonzero branching ratio for $\tau\rightarrow \mu ee$ ) to
get the desired $\nu_{\tau}$ lifetime of order 100 sec.

Another point worth noting is that unlike
 the singlet majoron model, the lifetime of $\nu_{\tau}$
in this case is independent of the $\nu_e$ mass. Therefore, this model will
remain viable even if the upper limits on the $\nu_e$ mass go down to the level
of $.1$ eV from future neutrinoless double beta decay searches.

\vspace{6mm}

\noindent{\bf V. Conclusion:}

\vspace{6mm}

We have pointed out two astrophysical constraints on an MeV range
tau neutrino: the first one has to do with an unconventional MeV
$\nu_{\tau}$ which carries a small electric charge that could make
a potential dark matter candidate. Our investigation shows that this
interesting possibility runs into conflict with the data on 511 keV
gamma rays obtained with the help of the Gamma Ray Observatory.
The second one deals with the conventional electrically neutral
$\nu_{\tau}$ with an MeV range mass, which is unstable. We show that
if decays to $\nu_e~+~\chi$, then SN1987A neutrino observations
imply an upper as well as lower bound on its lifetime. This allowed
window is interesting in the sense that it coincides with one considered
desirable by Dodelson et al.\cite{dgt}
 in their attempts to understand structure
in the universe using only cold dark matter plus a decaying Mev range
tau neutrino. We then discuss several gauge models where such lifetimes
can arise naturally in the hope that this new scenario for structure
formation will receive more serious consideration by both cosmologists
as well as particle physicists.

\vspace{6mm}
\begin{center}
{\bf Acknowledgement}
\end{center}
\vspace{6mm}

One of the authors (S. N.) would like to thank R. Ramaty for
discussions and the Nuclear Theory group of the University of
Maryland for kind hospitality.

\end{document}